\documentclass[%
 reprint,
 amsmath,amssymb,
 aps,
]{revtex4-2}

\usepackage{graphicx}% Include figure files
\usepackage{dcolumn}% Align table columns on decimal point
\usepackage{bm}% bold math
\usepackage{xcolor}% for text color
\begin{document}

\preprint{APS/123-QED}

\title{Unifying Interpretations of Phase Transitions in the Vicsek Model: 
Correlation Length as a Diagnostic Tool}

\author{Wenhao Yu}

\author{Zinuo Li}

\author{Liufang Xu}%
\email {Corresponding email: {lfxuphy@jlu.edu.cn}}

\affiliation{Biophysics \& Complex System Center,
Center of Theoretical Physics, College of Physics,
Jilin University Changchun 130012,
People's Republic of China}

\date{\today}% It is always \today, today,
%%%%%%%%摘要%%%%%%%%%%
\begin{abstract}
Vicsek Model is widely used in simulations of dry active matter. 
We re-examined two typical phase transitions in the original Vicsek model 
by using the velocity correlation length. One is the noise-driven 
disordered-to-ordered phase transition driven by noise, which was initially 
considered as a second-order transition (continuous transition), but was later 
demonstrated by Chate's detailed study to be a first-order transition. The other one 
is the disordered-to-ordered phase transition driven by average distance between particles, 
which is a second-order transition and satisfies the hyper-scaling relation of continuous 
transitions. We have discovered the change of correlation length during transition indicates 
a critical point in continuous transition while not in the discontinuous situation. 
We have also provided a method to classify phase transitions in active matter systems by 
using the correlation length and summarized previous work within the same framework. 
Finally, we end up with a potential application in experiments of bactirial swarms and 
robotic swarms. 
We hope our work paves the way for both theory and experiment development of active matter.

\end{abstract}

\maketitle

%\tableofcontents

\section{\label{sec:level1}Introduction}

 Active matter systems represent a class of non-equilibrium systems comprising 
energy-consuming self-propelled particles that generate mechanical motion through 
internal energy dissipation\cite{toner2005a,marchetti2013,ramaswamy2010}. 
Fascinating collective phenomena observed in bacterial
systems have been reported recently, including large-scale synchronized 
oscillations\cite{chen2017a}, emergent vortex patterns\cite{xu2024}. Collective 
behavior, which has aroused great research interest among relevant 
scientific researchers, extend across multiple scales, from macroscopic fish 
and starling birds\cite{ScalefreeCorrelationsStarling} to microscopic 
bacterial colonies\cite{attanasi2014,berg2000}. 
 The computational study of dry active matter has advanced significantly\cite{binder1997,kürsten2020,caussin2014}. 
since the seminal Vicsek model (VM)\cite{vicsek1995} demonstrated spontaneous 
symmetry breaking and a disorder-order phase transition in two-dimensional systems. 
 This minimal model exhibits the existence of long-range order
in a two dimension space and a disordered-to-ordered phase 
transition. However, the phase transition from disorder to c
ollective behavior in dry active matter remains a highly 
controversial topic\cite{martin2024a}. 
 Initial interpretations suggested a continuous second-order transition\cite{vicsek1995},
while subsequent studies employing finite-size scaling analysis of Binder cumulants 
revealed first-order characteristics\cite{grégoire2004}. This historical dichotomy 
motivates our re-examination through the fundamental perspective of spatial correlations. 
 Focusing on Z$_2$ symmetry breaking transitions, we establish correlation length 
divergence as a universal signature of continuous transitions, while demonstrating 
its absence in discontinuous counterparts. This methodology not only reconciles 
previous conflicting results but also provides a transferable framework for 
characterizing non-equilibrium phase transitions.

Recent studies have systematically investigated phase transitions 
in the Vicsek model (VM) and its modified variants. In extended systems, 
the flocking transition exhibited by the Vicsek model
demonstrates striking analogy to nonequilibrium fluctuation-induced 
liquid-gas transitions \cite{solon2015,martin2024}. Developments in 
chiral active matter systems have revealed PT symmetry breaking mechanisms 
through modified VM implementations, uncovering distinctive non-reciprocal 
phase transitions \cite{fruchart2021}. Despite these advances, the fundamental 
classification of phase transitions in active matter systems remains a 
persistent challenge, particularly regarding the precise characterization 
of discontinuous transition regimes.

Correlations constitute a fundamental mechanism underlying collective 
behavior \cite{jensen2021,cavagna2018,vicsek2012}. Long-range correlations 
reflect a system's capacity for group-wide information transfer, 
an essential prerequisite for coordinated responses \cite{ScalefreeCorrelationsStarling}. 
And it plays a role of a cornerstone in 
describing collective behavior in active matter and flocks such as starling 
flocks\cite{cavagna2018} from the sight of statistical mechanics. Simulations and 
experiments have confirmed that a system would need to readjust its control
parameter according to its size to be maximally correlated\cite{attanasi2014a}, which 
indicates a critical point in change of correlation length. We believe 
that this change is related to the critical phase transition that the 
system undergoes and few studies on judging the type of phase transition 
in a VM by the correlation length are reported. Addressing 
this theoretical-experimental gap, we establish correlation 
length scaling as a unifying framework for resolving phase 
transition classification controversies.

In this paper, We present numerical simulations of two distinct disorder-to-order 
phase transitions in the VM. The first constitutes 
a continuous transition driven by density variations through 
inter-agent distance modulation\cite{ScalefreeCorrelationsStarling}, 
maintaining constant particle numbers. Through finite-size scaling 
analysis of the order parameter and its dynamic fluctuations, 
we obtained critical exponents and revealed previously 
underexplored correlation length evolution across the transition.
 The second involves noise-induced discontinuous transitions. 
Applying the same analytical framework, we observed exceptional 
behavior: Correlation length variations lack critical divergence 
characteristics, providing diagnostic power to differentiate 
transition types. Our comparative analysis elucidates fundamental 
differences in correlation length dynamics between continuous 
and discontinuous phase transitions, proposing mechanistic 
explanations for these contrasting behaviors.

The paper is organized as follows: Section II presents the Vicsek 
model (VM) implementation and provides a concise overview of 
correlation length theory and finite-size scaling in continuous 
phase transitions. Section III details our comparative analysis 
of disorder-to-order transitions through the correlation length 
framework, integrating previous theoretical approaches. Section V 
synthesizes key findings and discusses implications for phase 
transition classification in active matter systems.

\section{Model}
\subsection{\label{sec:citeref}Vicsek Model}

In this section, we will present the standard formulation of the Vicsek model (VM),  
a paradigmatic active matter system exhibiting collective motion. 
The VM describes the collective dynamics of self-propelled particles
interacting through local alignment rules. 
Self-propelled particles move at constant speed $v_0$ while aligning  
their velocities with neighbors within a fixed interaction radius $R_c$.  
The discrete-time dynamics for particle $i$ with position $\bm{r}_i$ and  
orientation $\theta_i$ evolves as:

\begin{equation}
\bm{r}_i(t+\Delta t) = \bm{r}_i(t) + v_0 \Delta t \bm{n}_i(t)
\end{equation}

\begin{equation}
\theta_i(t+\Delta t) = \langle \theta_i(t) \rangle_{R_c} + \eta \xi(t)
\end{equation}

where $\bm{n}_i = (\cos\theta_i, \sin\theta_i)$ is the orientation vector,  
$\xi(t) \in [-\pi,\pi]$ represents uniform angular noise, and $\eta$  
controls noise intensity. The alignment interaction computes the  
local average orientation through:

\begin{equation}
\langle \theta_i(t) \rangle_{R_c} = \mathrm{Arg}\left( \sum_{j \in \mathcal{N}_i} 
e^{i\theta_j} \right)
\end{equation}

with $\mathcal{N}_i = \{ j : \|\bm{r}_j - \bm{r}_i\| \leq R_c \}$.  
We implement $\Delta t = 1$ and $v_0 = 0.03$ throughout simulations,  
with system size $L$ and density $\rho = N/L^2$ as control parameters.  
The angular noise is scaled to $[-\eta/2, \eta/2]$ through parameter $\eta$.

The VM exhibits  disorder-order phase transition,  
quantified by the polarization order parameter:

\begin{equation}
\phi = \frac{1}{Nv_0} \left| \sum_{i=1}^N \bm{v}_i \right| 
\end{equation}

where $\bm{v}_i = v_0\bm{n}_i$. This parameter satisfies $\phi \to 1$  
for perfect alignment and $\phi \to 0$ in disordered states.  
Critical fluctuations are captured by the susceptibility:

\begin{equation}
\chi = N \left[ \langle \phi^2 \rangle - \langle \phi \rangle^2 \right]
\end{equation}

where $\langle \cdot \rangle$ denotes temporal averaging after  
thermalization. The peak position of $\chi$ identifies the critical point.

\subsection{\label{sec:citeref}Spatial velocity correlations and correlation length}

We introduce spatial velocity correlations as an alternative approach  
to characterize phase transition nature. Both ordered and disordered phases  
exhibit distinctive correlation patterns, revealing fundamental features  
of collective dynamics. The equal-time connected velocity correlation  
function is defined as \cite{cavagna2018}:

\begin{equation}
C(r) = \frac{\sum_{i<j}^N \delta(r - r_{ij}) \delta\bm{v}_i \cdot \delta\bm{v}_j}
{\sum_{i<j}^N \delta(r - r_{ij})}
\end{equation}

where $r_{ij} \equiv \|\bm{r}_i - \bm{r}_j\|$ denotes interparticle distance,  
$\delta\bm{v}_i \equiv \bm{v}_i - \langle\bm{v}\rangle$ represents velocity  
fluctuation from the global average $\langle\bm{v}\rangle = \frac{1}{N}\sum_i \bm{v}_i$,  
and $\delta(\cdot)$ is the Dirac delta function. The numerator accumulates  
correlated velocity fluctuations at separation $r$, normalized by the  
corresponding particle pair count in the denominator.

The correlation length $\xi$ is operationally defined as the smallest  
$r$ satisfying $C(r) = 0$, marking the crossover from positive correlations  
($C(r) > 0$) to anticorrelations ($C(r) < 0$). This threshold identifies  
the characteristic domain size of coherent motion. While $C(r)$ may  
possess multiple zeros, we systematically select the first root as $\xi$.  

At continuous phase transitions, $\xi$ diverges algebraically near the  
critical point. This divergence behavior provides crucial insights  
into the universality class of the transition. Our finite-size scaling  
analysis utilizes $\xi$ as a key diagnostic for distinguishing between  
first-order and continuous phase transitions.

\subsection{\label{sec:citeref}Finite size scaling and continuous transition}
Phase transitions can be classified as continuous (second order) and discontinuous (first order) 
phase transitions. For continuous phase transitions, critical points and critical characteristics can 
be observed from numerical simulations. From decades of research in critical phenomena, it has established
 that continuous transitions exhibit rounding and shifting effects in critical point that can be accounted 
 for by means of the standard finite-size scaling theory\cite{Cardy1996,Baglietto2008}. Within this framework, the scaling ansatz for 
 the order parameter of the VM can be written as
\begin{equation}
\phi(N,d)=N^{-\beta/\nu}f((d-d_c)L^{1/2\nu})
\end{equation}
where $\beta$ and $\nu$ are the critical exponents of the order parameter and correlation length, respectively,
and $f$ is a scaling function. The critical point $\eta_c$ is the noise strength at which the phase transition
occurs in the thermodynamic limit. The scaling function $f$ is expected to be a universal function that depends
only on the scaling variable $(\eta-\eta_c)L^{1/\nu}$, and the critical exponents $\beta$ and $\nu$ are universal
quantities that characterize the critical behavior of the system. The critical exponents $\beta$ and $\nu$ are
related to the scaling of the order parameter and correlation length, respectively, and they are expected to be
the same for all systems in the same universality class. The finite-size scaling theory predicts that the order
parameter $\phi$ scales with the system size $L$ as $L^{-\beta/\nu}$ at the critical point $\eta_c$.  Also, the 
finite-size scaling ansatz for $\chi$ reads 
\begin{equation}
\chi(N,d)=N^{\gamma/\nu}g((d-d_c)d^{1/2\nu})
\end{equation}
where $\gamma$ is the critical exponent of the susceptibility $\chi$. The scaling function $g$ is expected to be
a universal function that depends only on the scaling variable $(\eta-\eta_c)L^{1/\nu}$, and the critical exponent
$\gamma$ is a universal quantity that characterizes the critical behavior of the system. For standard critical 
phenomena, we have the hyper-scaling relationship
\begin{equation}
2\beta+\gamma=d\nu
\end{equation}
where $d$ is the dimension of the system. The hyper-scaling relationship is a consequence of the scaling ansatz
which is expected to hold for continuous phase transitions. The critical exponents $\beta$, $\gamma$, and $\nu$
are expected to satisfy the hyper-scaling relation for continuous phase transitions. The critical exponents $\beta$,
$\gamma$, and $\nu$ are expected to be the same for all systems in the same universality class. we will attempt 
to determine relevant critical exponents and test the validity of the hyper-scaling relationship in the 
below sections

\section{Results and Discussion}
\subsection{\label{sec:citeref}Density-induced phase transition and correlation length}

We first investigate an unconventional continuous phase transition
in active matter systems. Simulations reveal that dry active matter
systems exhibit maximal correlation length when control parameters
are adjusted relative to system size\cite{grégoire2004}. This divergence signals
critical behavior at the order-disorder transition point, confirming
the transition's fundamental nature.

By fixing noise strength $\eta=2$ while systematically varying
particle density $\rho = N/L^2$ (equivalent to tuning average
interparticle distance $d=1/\sqrt{\rho}$), we observe a distinct
density-induced phase transition in the Vicsek model (VM).
Numerical simulations with $N=\{200,400,800,1600\}$ and $L\in[0.1,3.5]$
reveal characteristic critical signatures. Figure 1(a) demonstrates
the order parameter $\phi$ evolution, while Fig. 1(b) shows
corresponding dynamic fluctuations $\chi$, both exhibiting
hallmarks of continuous transitions. 

\begin{figure}[h]
  \includegraphics[width=\linewidth]{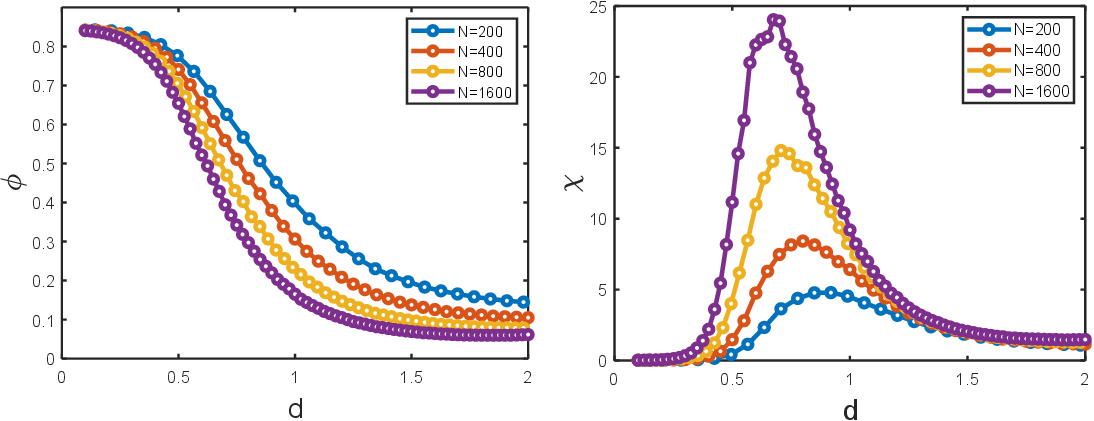}% Resize figure to fit within page width
    \caption{\label{fig:epsart1} (a) Order parameter $\phi$ as a 
    function of average distance $d$ for different system sizes $L$. 
    (b) Dynamic fluctuations $\chi$ as a function of average distance 
    $d$ for different system sizes $L$. }
\end{figure}

Finite-size scaling analysis confirms critical scaling behavior.
As shown in Fig. 2(a-b), log-log plots of $\phi_c(N)$ and $\chi_c(N)$
yield critical exponent ratios $\beta/\nu=0.141(46)$ and
$\gamma/\nu=1.562(41)$. These satisfy the hyperscaling relation
$2\beta/\nu+\gamma/\nu=1.844(133)$, compatible with $d=2$ dimensions
within error margins. This consistency validates the continuous
transition framework for density-driven criticality.

\begin{figure}[h]
  \includegraphics[width=\linewidth]{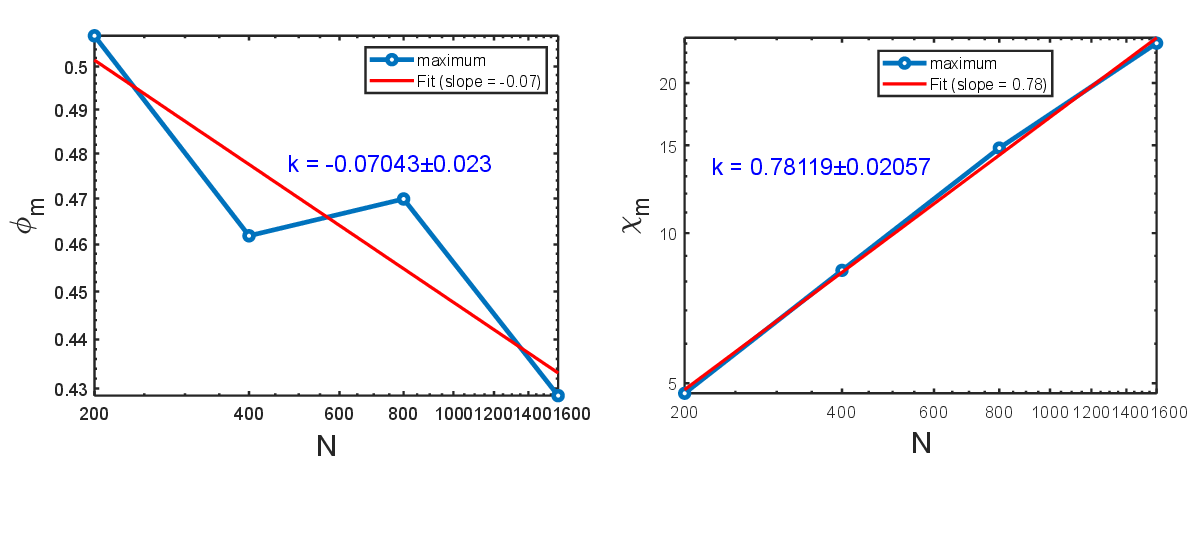}% Resize figure to fit within page width
    \caption{\label{fig:epsart2} (a) Log-log plot of the order parameter $\phi$ at 
    the critical point as a function of the number of particles $N$.
    The slope of the line gives the critical exponent ratio $\beta/\nu$. 
    (b) Log-log plot of the dynamic fluctuations $\chi$ at the critical point 
    as a function of the number of particles $N$. The slope of the line gives 
    the critical exponent ratio $\gamma/\nu$.}
  \end{figure}

It is worth noting that, in this phase transition, the change of correlation 
length is also similar to a traditional continuous phase transition, with 
a maximum value at the critical point, as shown in Fig 3(a). The correlation 
length retains the same characteristics as dynamic fluctuations and shows the 
possible presence of a critical point, which is characteristic of 
continuous phase transitions. The results of the correlation length
are consistent with the critical exponents we calculated above. And the
correlation length at the critical point is proportional to the number of particles
$N$, as shown in Fig 3(b), indicates a power-law scaling relationship
between the correlation length and the number of particles, which reads 
\begin{equation}
\xi(N,d)=N^{k_1}\tilde{\xi}((d-d_c)d^{k_2}). 
\end{equation}

where $k_1$ and $k_2$ are critical exponents of the correlation length.
We can get $k_1=0.36$ resulting from Fig3(b).
This behavior mirrors 3D VM results \cite{grégoire2004}, confirming
universal critical phenomena across dimensionalities. The combined
evidence of scaling collapse, hyperscaling satisfaction, and
divergent correlations establishes this as a genuine critical
transition.

\begin{figure}[h]
\includegraphics[width=\linewidth]{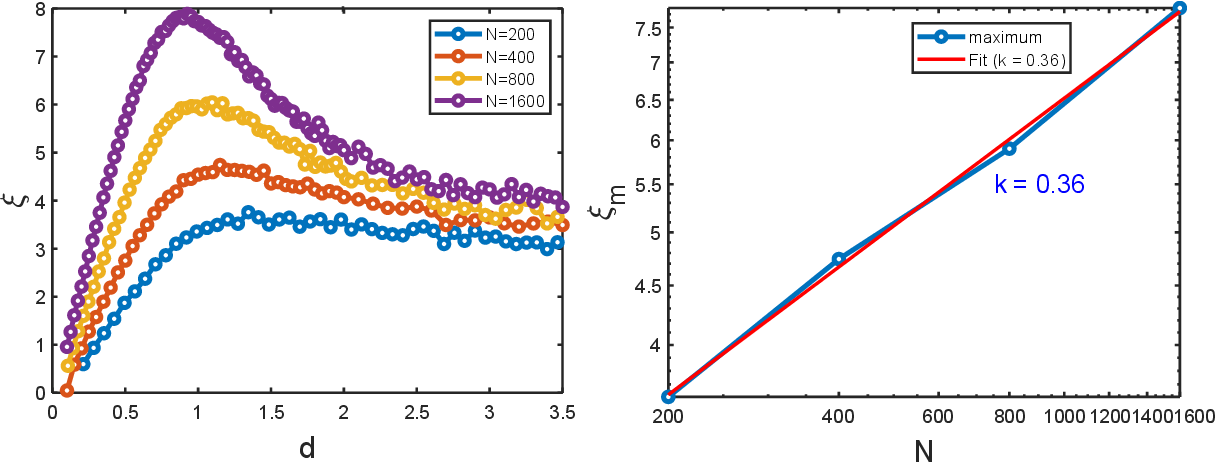}% Resize figure to fit within page width
  \caption{\label{fig:epsart2} (a) Correlation length $\xi$ as a
  function of average distance $d$ for particle numbers $N$. (b)log-log plot of the
  correlation length $\xi$ at the critical point as a function of
  the number of particles $N$. In each simulation. N is fixed and average distance
  between particles $d$ is changed.}
\end{figure}

%%%%%%%%%%%%%%%%%%%%%%%%%%%%%%%%%%%%%%%%%%%%%%%%%%%%%%%%%%%%%%%
\subsection{Noise-induced phase transition and correlation length}

The putative criticality in noise-induced VM transitions requires 
re-examination through correlation analysis. Our systematic study 
reveals fundamental differences from standard critical phenomena. 
Simulations employ $v_0=0.03$ with fixed density $\rho=4$ across 
system sizes $L=\{4,8,12,16,20\}$ (corresponding to $N=64$--$1600$),
employing $2\times10^6$ total samples (2000 steps/run $\times$ 1000 
realizations) to ensure statistical reliability.

Figure 4(a-b) demonstrates apparent critical signatures through 
$\phi(\eta)$ shifting and $\chi(\eta)$ peaking -- features 
originally interpreted as continuous transition evidence \cite{vicsek1995}.
It seems that the phase transition
is continuous only by studying the order parameter and its dynamic fluctuations.
In \cite{grégoire2004}, the Binder cumulant was used to determine the order of
phase transition, and the results showed that the transition was discontinuous. 
To better understand the nature of the phase transition its 
cirtical parts, we choose to study the correlation length as in the previous section.

\begin{figure}[h]
  \includegraphics[width=\linewidth]{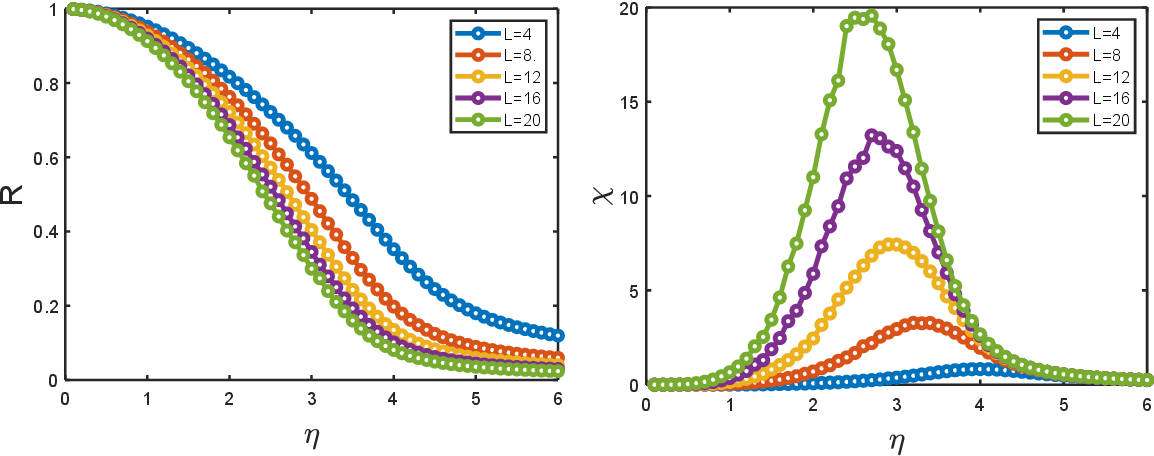}% Resize figure to fit within page width
    \caption{\label{fig:epsart3}(a) Order parameter $\phi$ as a 
    function of noise strength $\eta$ for different system sizes $L$. 
    (b) Dynamic fluctuations $\chi$ as a function of noise strength $\eta$
     for different system sizes $L$.}
  \end{figure}

However, the relationship between $\xi$ and $\eta$ in Fig.4(a) seems not to support 
the view above. In systems with different size, the changes of correlation 
length are similar. In low noise area and a small system size, the 
correlation length is relatively flat or even keep a fixed value, which 
indicates that the system has a certain resistance to noise. At the same 
time, the flat area also illustrates the existence of coexistence states. 
With the noise strength crossing the critical value $\eta_c$, the correlation 
length has suddenly decreasesd. We didn't observe the similar trend in 
correlation length as we talked in III.A. In place of the critical point 
is a relatively flat maximum of the correlation length over a wide range 
of noise strength. Compared with the correlation length in continuous 
transition, the characteristics of second-order phase transitions are 
not reflected here. Due to the singular variation of $\xi$, the transition 
is discontinuous as a result.

\begin{figure}[h]
  \includegraphics[width=\linewidth]{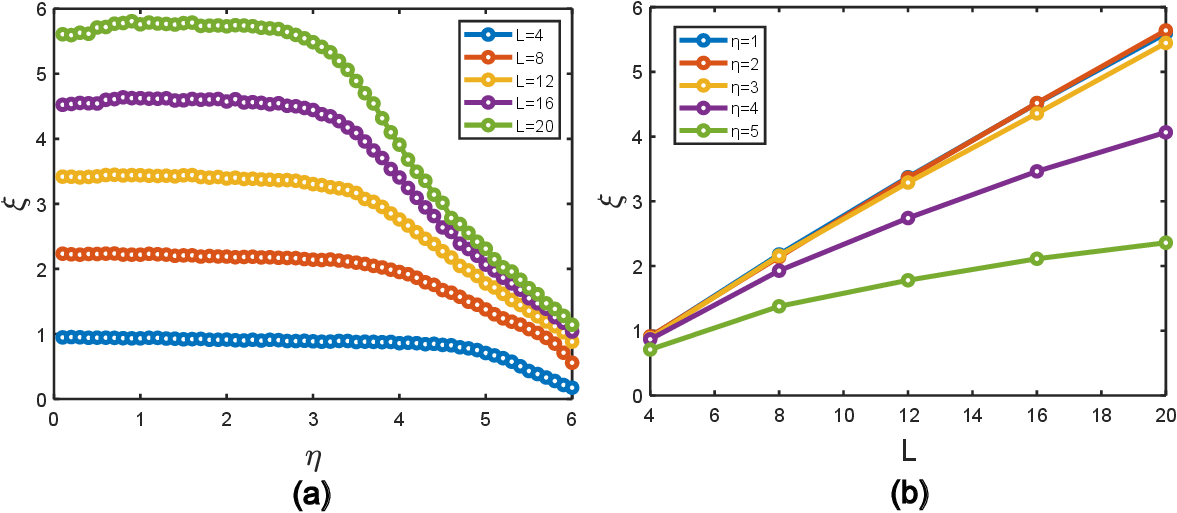}% Resize figure to fit within page width
    \caption{\label{fig:epsart3}(a) Correlation length $\xi$ as a
    function of noise strength $\eta$ for different system sizes $L$.
    (b) The dependence of correlation length $\xi$ on L in different fixed noise 
    strength $\eta$. Here we choose parameter $\rho=4$ with L and $\eta$ changing.}
\end{figure}

Though in the low noise region and low system size, we have 
$\xi(\eta_x)\propto L$, consistent with the results in\cite{cavagna2018}. 
However, this linear relationship is broken with noise strength increasesing. 
Fig 4(b) shows the dependence of correlation length on system size under 
different noise strength. In relatively high noise strength region ($\eta=5$), 
the proportional relationship between $\xi$ and $L$ is broken. 
Furthermore, it is also important to  notice that the relatively flat section of the correlation 
length becomes shorter as the size of system increases. As the system size 
reaches infinite, the criticality and scale-free properties showed in Fig 
4(b) will disappear, which indicates that the origin of critical properties 
and initial thought of continuous transition is finite size effect.

The anomalous results of $\xi$ against $\eta$ in the low noise region is also
observed in confined active brownian particles\cite{caprini2021} while firstly
reported in our work for VM. It suggests universal mechanical stabilization 
in correlations between particles against weak perturbations.

\subsection{Absence of criticality in the infinite system size limit}

To better illustrate the above points, we focus on the finite-size effect 
by fixing total number of particles N=512 while varying system size L. Fig 5(a) 
shows how system size effect the trend of correlation length against noise strength. 
As predictions, the correlation length in a small system size is relatively flat
in a wide range of noise strength, which indicates the existence of resistance to
noise and coexistence states. The correlation length in a large system size is more sensitive 
to noise strength, which indicates the system is more likely to be in a disordered state.
Fig 5(b) shows the correlation length against system size for different fixed noise strength.
The antatz $\xi\sim L$ is valid in a low noise region, which is consistent with the
results in \cite{cavagna2018,attanasi2014a}. However, this linear relationship is 
totally disappear when reaching relative large systems even in a low noise region.
The ralatively flat section disappears as the system size
increases, which indicates that the criticality and scale-free properties showed
in Fig 4(b) will disappear as the system size reaches infinite. The results
indicate that the origin of critical properties and initial thought of continuous
transition is finite size effect.

\begin{figure}[h]
\includegraphics[width=\linewidth]{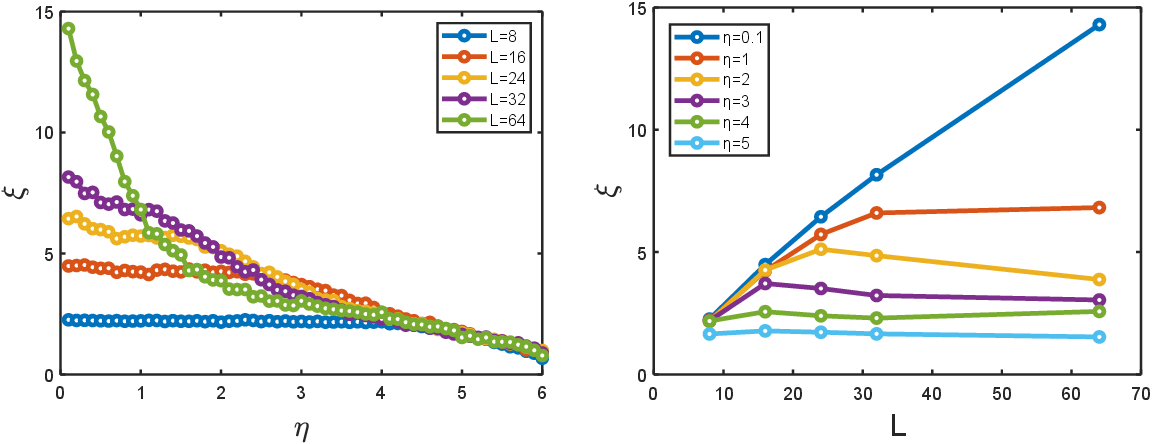}
\caption{\label{fig:epsart3}(a)Correlation length $\xi$ against
noise strength $\eta$ for different system sizes $L$ with fixed number of 
particles $N=512$.(b)$\xi$ against $L$ for different fixed noise strength $\eta$
and collapse of criticality.}
\end{figure}

\subsection{Potential applications in experiments}

The correlation length method proposed in this paper 
offers versatile tools for probing collective dynamics 
in both biological and engineered systems. Here, we elaborate 
on its potential applications in experimental studies of 
bacterial swarms and robotic swarms, highlighting how this 
approach can uncover fundamental insights into phase transitions 
and emergent order.

\textbf{Bacterial swarms.} In biological systems such as 
bacterial colonies (e.g., \textit{Bacillus subtilis} or \textit{E. coli}), 
the correlation length method can quantify collective motion and 
pattern formation during swarming. By analyzing time-resolved 
imaging data—such as velocity fields derived from particle image 
velocimetry or tracked trajectories of fluorescently labeled 
bacteria—the correlation length can be measured to determine the 
spatial scale over which individual motions are correlated. This 
metric may identify critical thresholds in environmental parameters 
(e.g., nutrient concentration, shear stress, or cell density) that 
trigger transitions between disordered and ordered states. For instance, 
a divergence-like behavior in correlation length near a critical density would signal 
a continuous (second-order) phase transition, whereas abrupt jumps in 
correlation length might indicate discontinuous (first-order) transitions.
Such insights could clarify how bacterial collectives optimize survival 
strategies, such as biofilm formation or antibiotic evasion. Practically, 
this method could diagnose swarm states in real time, enabling interventions 
to disrupt pathogenic biofilms or enhance beneficial microbial communities.

\textbf{Robotic swarms.} In engineered systems like 
robotic swarms (e.g., Kilobots or drone fleets), the correlation length
method provides a quantitative framework to optimize interaction rules
and detect emergent pathologies. By programming robots with tunable 
communication ranges or alignment strengths, experiments could measure 
how correlation length varies with these parameters using positional 
or velocity data. For example, increasing the interaction radius might 
extend correlation length, fostering global synchronization, while noise
or limited bandwidth could fragment correlations, signaling swarm 
instability. This approach allows rigorous testing of theoretical 
models, such as Vicsek-like flocking, and could guide the design of 
adaptive algorithms that balance cohesion and energy efficiency. In
applications like search-and-rescue or environmental monitoring,
real-time correlation length metrics could ensure swarm robustness, 
triggering self-repair mechanisms if correlations drop below critical
thresholds. Furthermore, studying phase transitions in robotic systems
might inspire bio-inspired strategies, bridging insights from biological
and artificial swarms.

These potential applications underscore the method's 
broad utility in both deciphering natural phenomena and 
enhancing engineered systems, establishing correlation length 
as a universal metric for cross-disciplinary swarm studies.

\section{Conclusion}
Our investigation of the Vicsek model from the point of view of 
velocity correlation length resolves long-standing ambiguities in 
classifying phase transitions within active matter. By distinguishing 
between noise-driven and density-driven transitions, we state that the
correlation length $\xi$ serves as a universal criterion for identifying 
transition order.

For noise-driven transitions, our results align with Chate's conclusion 
when the system size is large the transition is discontinuous (first-order).
The absence of $\xi$ divergence at criticality, even in finite systems, and 
the disappearance of scale-free correlations in the thermodynamic limit 
(\(L \to \infty\)) confirm that earlier interpretations of continuity were 
artifacts of finite-size effects. This reconciles contradictions between 
initial claims of criticality and later evidence of phase coexistence. 
Notably, our method bypasses the need for direct measurement of dynamic 
hysteresis loops, offering a simpler geometric proxy—the first zero-crossing 
of \(C(r)\)—to detect discontinuities.

In contrast, density-driven transitions exhibit hallmark features of 
second-order criticality: $\xi$ diverges at the critical point, and hyper-scaling 
relations hold rigorously. The success of the hyper-scaling law (\(\gamma + 2\beta = \nu d\)) 
underscores the robustness of equilibrium critical phenomena frameworks in 
certain non-equilibrium active systems. This suggests that density-driven 
ordering in the Vicsek model belongs to a universality class governed by 
conserved dynamics, akin to equilibrium ferromagnetic transitions.

The stark contrast in $\xi$ behavior between these transitions 
highlights the dual role of interaction range and symmetry-breaking 
mechanisms in active matter. Noise-driven transitions, governed by 
alignment frustration and local fluctuations, suppress long-range 
order unless constrained by finite-size boundaries. Density-driven 
transitions, however, amplify collective alignment through proximity-enhanced 
interactions, enabling true criticality. Our findings further emphasize 
that finite-size effects, while often unavoidable in simulations, can mask 
intrinsic physics a cautionary note for interpreting scale-free correlations 
in bounded systems.

In conclusion, we examined phase transitions in the Vicsek model using velocity 
correlation length as a diagnostic tool. Key conclusions include Classification 
Framework, hyper-Scaling Validation and finite-Size Artifacts. Our 
correlation-length-based framework unifies disparate interpretations of 
Vicsek model transitions and provides a scalable methodology for probing phase behavior
in other active systems, such as motility-induced phase separation or chiral flocks. 
Future work could extend this approach to heterogeneous or 
anisotropic systems, explore dynamical correlation lengths, and test universality classes
across diverse active matter models. By bridging concepts from equilibrium statistical 
mechanics and non-equilibrium physics, this work advances the theoretical toolkit for 
understanding collective behavior in biological and synthetic active materials.

\begin{acknowledgments}
The authors thank Haoran Xu from Zhejiang University for many fruitful discussions.
W.Y. also thanks Zhengbang Hu, Mingseng Fan, Jiewen Lu and Jinni Yang from 
Jilin University.
\end{acknowledgments}

\nocite{*}

\bibliography{apssamp}% Produces the bibliography via BibTeX.

\end{document}